\documentclass[11pt,a4paper]{article}
\pdfoutput=1
\usepackage{jheppub}
\usepackage{tikz}
\usetikzlibrary{intersections, calc, arrows.meta, bending}
\usepackage{subcaption}
\DeclareMathOperator{\tr}{tr}

\newcommand{\ri}{\mathrm{i}}
\renewcommand{\th}{\theta}

\newcommand{\hf}{\frac{1}{2}}

\newcommand{\til}[1]{\widetilde{#1}}

\newcommand{\del}{\partial}

\newcommand{\bra}{\langle}
\newcommand{\ket}{\rangle}
\newcommand{\la}{\lambda}

\newcommand{\bt}{\beta}

\newcommand{\rt}[1]{\sqrt{#1}}
\newcommand{\cO}{\mathcal{O}}

\newcommand{\cF}{\mathcal{F}}

\makeatletter
\gdef\@fpheader{}
\makeatother

\begin{document}
\title{High temperature expansion of double scaled SYK}

\author{Kazumi Okuyama}

\affiliation{Department of Physics, 
Shinshu University, 3-1-1 Asahi, Matsumoto 390-8621, Japan}

\emailAdd{kazumi@azusa.shinshu-u.ac.jp}

\abstract{We study the high temperature (or small inverse temperature
$\beta$) 
expansion of the free energy
of double scaled SYK model.
We find that this expansion is a convergent series with a finite radius of convergence. It turns out that the radius of convergence
is determined by the first zero of the partition function on the imaginary
$\beta$-axis.
We also show that the semi-classical expansion of the free energy
obtained from the saddle point approximation of the exact result
is consistent 
with the high temperature expansion of the free energy.}

\maketitle

\section{Introduction}
The Sachdev-Ye-Kitaev (SYK) model is a very useful toy model for the
study of quantum gravity  
\cite{Sachdev1993,Kitaev1,Kitaev2,Polchinski:2016xgd,Maldacena:2016hyu}.
At low energy, the SYK model is described by the Schwarzian mode
and it is holographically dual to the Jackiw-Teitelboim gravity
\cite{Jackiw:1984je,Teitelboim:1983ux}.

One can go beyond the low energy limit
by taking a certain double scaling limit of the SYK model 
\cite{Cotler:2016fpe}, which we call the DSSYK model in this paper.
The Hamiltonian of the 
SYK model is given by the random $p$-body interaction 
of $N$ Majorana fermions and the DSSYK model is defined by the limit
\begin{equation}
\begin{aligned}
 N,p\to\infty\quad\text{with}\quad \la=\frac{2p^2}{N}:~\text{fixed}.
\end{aligned} 
\end{equation}
It turns out that the partition function and the correlation functions
of DSSYK model reduce to the computation of the intersection numbers of 
chord diagrams.
For instance, the partition function is schematically written as
\begin{equation}
\begin{aligned}
 Z(\bt)=\bra\tr e^{-\bt H}\ket_J=\sum_{n=0}^\infty\frac{\bt^{2n}}{(2n)!}
\sum_{\text{chord diagrams}}q^{\# \text{intersections}},
\end{aligned} 
\end{equation}
where $\bra\cdots\ket_J$ represents the average over the random coupling $J$
and 
\begin{equation}
\begin{aligned}
 q=e^{-\la}.
\end{aligned} 
\label{eq:q-def}
\end{equation}
This counting problem of chord diagrams 
is exactly solvable using the technique of the transfer matrix 
\cite{Berkooz:2018jqr}.
One can take various limits of the parameters such as $\la$ and $\bt$
to study the bulk dual of DSSYK.
For instance, the semi-classical, small $\la$ limit of DSSYK
was recently considered in \cite{Goel:2023svz}. 
\footnote{
See also \cite{Lin:2022rbf,Berkooz:2022mfk,Okuyama:2022szh,Mukhametzhanov:2023tcg,Okuyama:2023bch} for recent developments in DSSYK.
}

In this paper, we will study the high temperature
(or small $\bt$) expansion of the free energy $\log Z(\bt)$ 
of DSSYK.
We find that this expansion is a convergent series with a 
finite radius of convergence, and the radius of convergence is determined
by the zero of the partition function along the imaginary $\bt$-axis.
We also compute the semi-classical, small $\la$ expansion
of the free energy up to $\cO(\la^2)$. We find that 
the small $\bt$ limit of this semi-classical expansion 
agrees with the small $\la$ limit of the high temperature expansion.
In other words, there is no order-of-limit problem between
the small $\la$ and the small $\bt$ expansions.

This paper is organized as follows.
In section \ref{sec:small}, we study the high temperature expansion of the free 
energy at fixed $q$. We find that this expansion is a 
convergent series with a 
finite radius of convergence.
In section \ref{sec:zero},
we find numerical evidence that the radius of convergence
of the high temperature expansion is related to the first zero of the
partition function along the imaginary $\bt$-axis.
 In section \ref{sec:pade}, we consider the Pad\'{e}
approximation of the high temperature expansion.
We find a good agreement between the Pad\'{e}
approximation and the exact result, even at large $\bt$.
This suggests that there is no Hawking-Page transition
for the partition function of DSSYK,
and the high and the low temperature regimes are smoothly connected.
In section \ref{sec:semi}, 
we compute the small $\la$ expansion of free energy at fixed $\bt$
up to $\cO(\la^2)$.
It turns out that this expansion is consistent with the small $\bt$
expansion of free energy at fixed $q$.
Finally, we conclude in section \ref{sec:conclusion}
with some discussion for future directions.

\section{Small $\bt$ expansion of free energy}\label{sec:small}
As shown in \cite{Berkooz:2018jqr}, the partition function of DSSYK
is given by
\begin{equation}
\begin{aligned}
 Z(\bt)=\bra 0|e^{\bt T}|0\ket,
\end{aligned} 
\label{eq:Zexact}
\end{equation}
where the transfer matrix $T$ is written 
in terms of the $q$-deformed oscillator $A_\pm$
\begin{equation}
\begin{aligned}
 T=A_{-}+A_{+}.
\end{aligned} 
\label{eq:T-mat}
\end{equation}
$A_{\pm}$ act on the chord number state $|n\ket$ and
they create or annihilate the chords
\begin{equation}
\begin{aligned}
 A_{-}|n\ket=\rt{\frac{1-q^n}{1-q}}|n-1\ket,\quad
A_{+}|n\ket=\rt{\frac{1-q^{n+1}}{1-q}}|n+1\ket.
\end{aligned} 
\end{equation} 
As shown in \cite{Berkooz:2018jqr},
$T$ can be diagonalized by the $q$-Hermite polynomial, from which one can derive
the integral representation of the partition function
\begin{equation}
\begin{aligned}
 Z(\bt)=\int_0^\pi\frac{d\th}{2\pi}(q;q)_\infty (e^{2\ri\th};q)_\infty
(e^{-2\ri\th};q)_\infty \exp\left(\frac{2\bt\cos\th}{\rt{1-q}}\right),
\end{aligned} 
\label{eq:Z-int}
\end{equation} 
where $(a;q)_\infty$ denotes the $q$-Pochhammer symbol
\begin{equation}
\begin{aligned}
 (a;q)_\infty=\prod_{n=0}^\infty (1-aq^n).
\end{aligned} 
\end{equation}
Note that $Z(\bt)$ is an even function of $\bt$
\begin{equation}
\begin{aligned}
 Z(-\bt)=Z(\bt).
\end{aligned} 
\end{equation}
 
From \eqref{eq:Zexact}, one can see that the partition function is
obtained once we know the moment $\bra 0|T^{2n}|0\ket$ of the transfer matrix $T$
\begin{equation}
\begin{aligned}
 Z(\bt)=\sum_{n=0}^\infty \frac{\bt^{2n}}{(2n)!}\bra 0|T^{2n}|0\ket.
\end{aligned} 
\end{equation}
The moment $\bra 0|T^{2n}|0\ket$ enumerates 
the intersection numbers of the 
chord diagram, whose explicit form
is known as the Touchard--Riordan formula \cite{touchard,riordan}
\begin{equation}
\begin{aligned}
 \bra 0|T^{2n}|0\ket&=\frac{1}{(1-q)^n}
\sum_{j=-n}^n(-1)^jq^{\hf j(j-1)}\binom{2n}{n+j}.
\end{aligned} 
\label{eq:moment}
\end{equation}

We are interested in small $\bt$ expansion of the free energy of DSSYK
at fixed $q$
\begin{equation}
\begin{aligned}
 \log Z(\bt)=\sum_{n=1}^\infty \frac{\bt^{2n}}{(2n)!}k_{2n}(q).
\end{aligned} 
\label{eq:free1}
\end{equation}
This expansion was considered in \cite{josuat2013cumulants}
and the coefficients $k_{2n}(q)$ were called cumulants in  
\cite{josuat2013cumulants}.
As noticed in \cite{josuat2013cumulants}, $k_{2n}(q)$
can be factorized as
\begin{equation}
\begin{aligned}
 k_{2n}(q)=(q-1)^{n-1}\til{k}_{2n}(q),
\end{aligned} 
\end{equation}
where $\til{k}_{2n}(q)$ is a polynomial in $q$ with degree
$\hf (n-1)(n-2)$.
The first few terms of $\til{k}_{2n}(q)$ read
\begin{equation}
\begin{aligned}
 \til{k}_2(q)&=1,\\
\til{k}_4(q)&=1,\\
\til{k}_6(q)&=q+5,\\
\til{k}_8(q)&=q^3+7 q^2+28 q+56,\\
\til{k}_{10}(q)&=q^6+9 q^5+45 q^4+165 q^3+450 q^2+918 q+1092.
\end{aligned} 
\end{equation}
As far as we know, the closed form of $\til{k}_{2n}(q)$ 
is not known in the literature.

\begin{figure}[t]
\centering
\includegraphics
[width=0.8\linewidth]{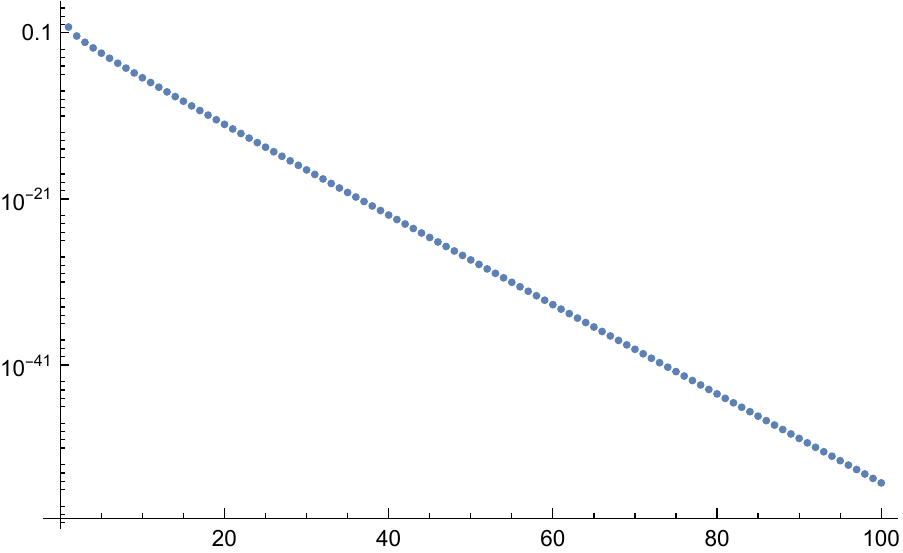}
  \caption{
Log-plot 
of $f_n(q)$ for $q=0.2$. The horizontal axis is $n$.
}
  \label{fig:logfn}
\end{figure}

\begin{figure}[t]
\centering
\includegraphics
[width=0.8\linewidth]{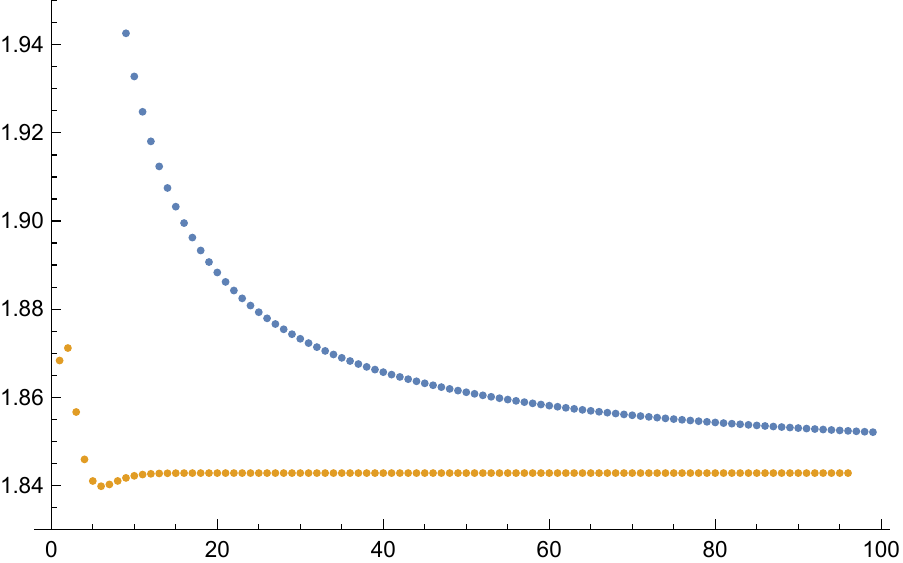}
  \caption{
Plot 
of $A_n(q)$ (blue dots) and $A_n^{(3)}(q)$ (orange dots)
for $q=0.2$. The horizontal axis is $n$.
}
  \label{fig:A-ric}
\end{figure}

\begin{figure}[t]
\centering
\includegraphics
[width=0.8\linewidth]{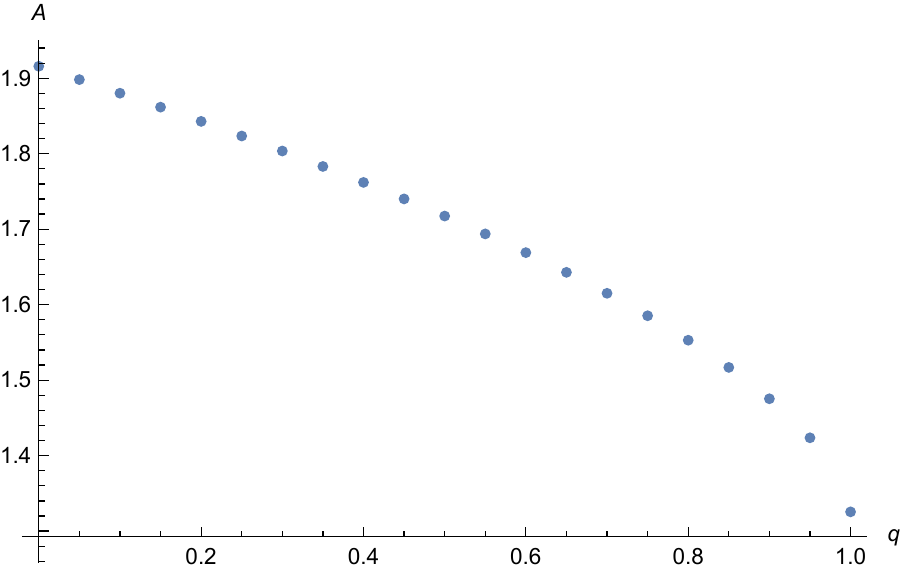}
  \caption{
Plot 
of $A(q)$ as a function of $q$.
}
  \label{fig:A}
\end{figure}

For our purpose, it is convenient to define $f_n(q)$ as
\begin{equation}
\begin{aligned}
 f_n(q)=\frac{\til{k}_{2n}(q)}{(2n)!}.
\end{aligned} 
\end{equation}
Then the small $\bt$
expansion of the free energy \eqref{eq:free1} becomes
\begin{equation}
\begin{aligned}
 \log Z(\bt)=\sum_{n=1}^\infty (-1)^{n-1}(1-q)^{n-1}\bt^{2n}f_n(q).
\end{aligned} 
\label{eq:free2}
\end{equation}
From the known formula of the moment \eqref{eq:moment}, one can easily
compute $f_n(q)$ up to very high order. 
We have computed $f_n(q)$ up to $n=100$ and we find numerically
that $f_n(q)$ decays exponentially at large $n$
\begin{equation}
\begin{aligned}
 f_n(q)\sim \frac{c(q)}{n}A(q)^{-2n},\qquad(n\gg1).
\end{aligned} 
\label{eq:fn-asy}
\end{equation}
See Figure \ref{fig:logfn} for the plot of $f_n(q)$ with $q=0.2$ as
an example.
One can extract $A(q)$ in \eqref{eq:fn-asy}
as the limit of the following sequence
\begin{equation}
\begin{aligned}
 A_n(q)=\rt{\frac{f_n(q)}{f_{n+1}(q)}}~~~ \xrightarrow{n\to\infty}~~~ A(q).
\end{aligned} 
\end{equation}
One can accelerate the convergence of the sequence by using
the technique of the Richardson extrapolation, where the $m$-th Richardson transform of the series 
$A_n(q)$ is defined by 
\footnote{See e.g. \cite{Marino:2007te} for a review of this method.}
\begin{equation}
\begin{aligned}
 A_n^{(m)}(q)=\sum_{k=0}^m(-1)^{k+m}\frac{(n+k)^m}{k!(m-k)!} A_{n+k}(q).
\end{aligned} 
\end{equation}
It turns out that $A_n^{(m)}(q)$ has a much faster convergence to $A(q)$
than the original sequence $A_n(q)$.
As an example, in Figure \ref{fig:A-ric}
we show the plot of $A_n(q)$ and its third Richardson transform
$A_n^{(3)}(q)$ for $q=0.2$. As we can see from
Figure \ref{fig:A-ric}, $A_n^{(3)}(q)$ converges to a constant much faster than
the original $A_n(q)$.
We have computed $A(q)$ numerically 
for various values of $q$ using the $20$-th Richardson transform.
See Figure \ref{fig:A} for the plot of $A(q)$ as a function of $q$.

We can also determine $c(q)$ in \eqref{eq:fn-asy}
as the limit of the following sequence
\begin{equation}
\begin{aligned}
 c_n(q)=nf_n(q)\left[\frac{nf_n(q)}{(n+1)f_{n+1}(q)}\right]^n~~~
\xrightarrow{n\to\infty}~~~c(q).
\end{aligned} 
\end{equation}
Again, using the technique of the Richardson extrapolation
we can accelerate the convergence of $c_n(q)$ to $c(q)$.
In this way, we find that $c(q)$ 
in \eqref{eq:fn-asy} has a simple form
\begin{equation}
\begin{aligned}
 c(q)=1-q.
\end{aligned} 
\end{equation}

\section{Interpretation of $A(q)$ as the first zero of $Z(\ri t)$}\label{sec:zero}
When $0\leq q\leq1$, the small $\bt$ expansion of the free energy
\eqref{eq:free2} is an alternating series
and the large $n$ asymptotics of $f_n(q)$ in
\eqref{eq:fn-asy} implies that this expansion is a convergent series 
with a finite radius of convergence
\begin{equation}
\begin{aligned}
 \log Z(\bt)&\approx \sum_{n\geq1}(-1)^{n-1}(1-q)^{n-1}\bt^{2n}\frac{1-q}{n}A(q)^{-2n}\\
&= \log\Bigl[1+\bt^2(1-q)A(q)^{-2}\Bigr].
\end{aligned} 
\label{eq:F-pole}
\end{equation} 
The radius of convergence is determined by the logarithmic singularity 
of \eqref{eq:F-pole}
at the imaginary value of $\bt$. If we define $\bt=\ri t$,
the singularity is located at $t=\pm t_1(q)$ with
\begin{equation}
\begin{aligned}
 t_1(q)= \frac{A(q)}{\rt{1-q}}.
\end{aligned} 
\label{eq:pole-t}
\end{equation}
As we will see below, this singularity corresponds to the first zero of the
partition function.

For instance, let us consider the $q=0$ case.
From the numerical analysis in the previous section, we find
that $t_1(q)$ at $q=0$ is estimated as
\begin{equation}
\begin{aligned}
 t_1(0)=A(0)=1.915852985103756157807\cdots.
\end{aligned} 
\label{eq:t0-num}
\end{equation}
When $q=0$, the measure factor in \eqref{eq:Z-int}
becomes a trigonometric function $(2\sin\th)^2$,
 and the exact partition function is simply given by
\begin{equation}
\begin{aligned}
 Z(\bt)=\int_0^\pi\frac{d\th}{2\pi}(2\sin\th)^2e^{2\bt\cos\th}=
\frac{1}{\bt}I_1(2\bt),
\end{aligned} 
\end{equation}
where $I_1(z)$ denotes the modified Bessel function of the first kind.
If we analytically continue $Z(\bt)$ to the pure imaginary $\bt=\ri t$, 
the partition function becomes
\begin{equation}
\begin{aligned}
 Z(\ri t)=\frac{1}{t}J_1(2t),
\end{aligned} 
\label{eq:ZJ}
\end{equation}
where $J_1(z)$ denotes the Bessel function.
\begin{figure}[t]
\centering
\includegraphics
[width=0.8\linewidth]{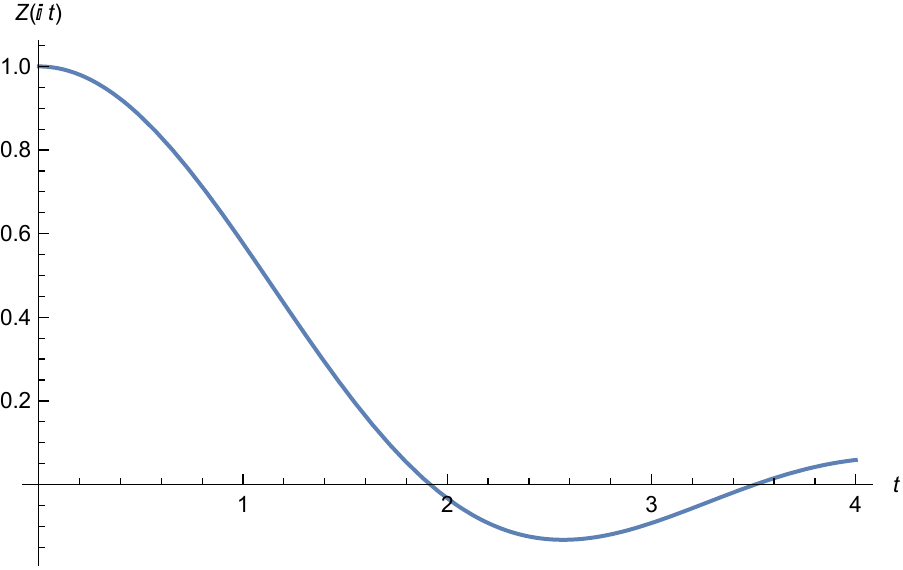}
  \caption{
Plot 
of $Z(\ri t)=\frac{1}{t}J_1(2t)$ for $q=0$.
}
  \label{fig:Z-q0}
\end{figure}
See Figure \ref{fig:Z-q0} for the plot of $Z(\ri t)$ in \eqref{eq:ZJ}.
$Z(\ri t)$ in \eqref{eq:ZJ} has zeros along the
real $t$-axis and the first zero on the positive $t$-axis is given by
\begin{equation}
\begin{aligned}
 t=\hf j_{1,1},
\end{aligned} 
\label{eq:j11}
\end{equation}
where $j_{n,k}$ is the $k$-th zero of the Bessel function $J_n(z)$.
In \texttt{Mathematica}, $j_{n,k}$ is implemented as
\texttt{BesselJZero[n,k]}.
Using this function in \texttt{Mathematica}, 
the numerical value of \eqref{eq:j11}
is evaluated as
\begin{equation}
\begin{aligned}
 \hf j_{1,1}=1.915852985103756157807\cdots,
\end{aligned} 
\end{equation}
which precisely matches \eqref{eq:t0-num}!
In fact, we found more than 20-digit agreement between 
$\hf j_{1,1}$ and $t_1(0)=A(0)$ obtained from the Richardson extrapolation.

\begin{figure}[t]
\centering
\includegraphics
[width=0.8\linewidth]{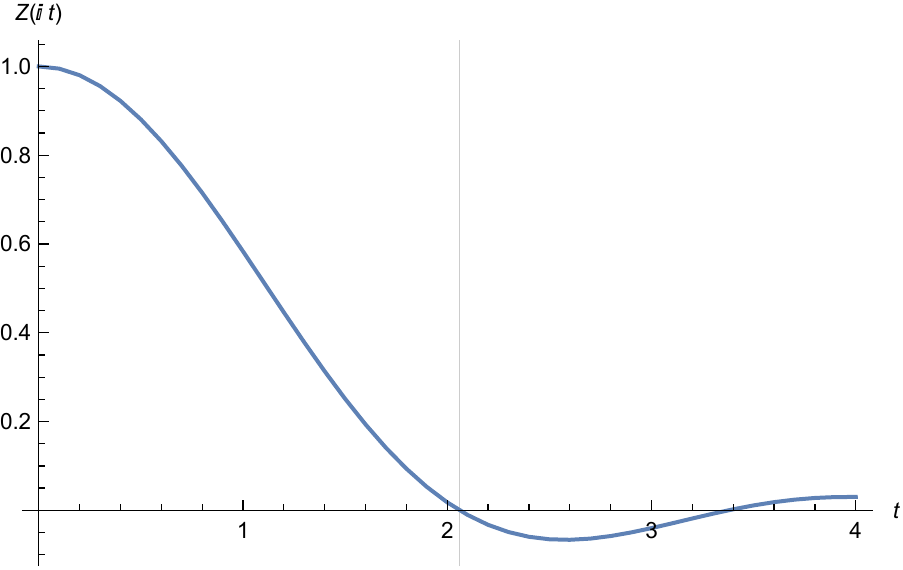}
  \caption{
Plot 
of $Z(\ri t)$ for $q=0.2$.
The gray vertical line is at $t=t_1(q)$, which
nicely agrees with the first zero of $Z(\ri t)$.
}
  \label{fig:Z-q02}
\end{figure}

For general $q$, we do not know the closed form
of the partition function $Z(\bt)$.
However, one can easily evaluate $Z(\bt)$
numerically using the integral representation of
$Z(\bt)$ in \eqref{eq:Z-int}.
As an example, in Figure \ref{fig:Z-q02}
we show the plot of $Z(\ri t)$ for $q=0.2$.
We indicated the location of $t=t_1(q)$ by the vertical gray line.
As one can see from Figure \ref{fig:Z-q02},
$t=t_1(q)$ obtained from the Richardson extrapolation nicely agrees with
the first zero of the partition function $Z(\ri t)$.
We have checked this agreement for several other values of $q$.

From the above numerical evidence, it is natural to conjecture that
the radius of convergence of the high temperature
expansion of the free energy is equal to the first zero
of the partition function $Z(\ri t)$ along the positive real $t$ axis.

$t_1(q)$ in \eqref{eq:pole-t} diverges as $q\to 1$ due to the denominator $\rt{1-q}$, 
but $A(q)$ remains finite at 
$q=1$. 
From the Richardson extrapolation, we find
\begin{equation}
\begin{aligned}
 A(1)=1.32548683869836316194948\cdots.
\end{aligned} 
\end{equation}
We observe that this agrees with the maximal value of the function $f(s)=2s/\cosh s$
\begin{equation}
\begin{aligned}
 \max_{s\in \mathbb{R}}\left\{\frac{2s}{\cosh s}\right\}=
1.32548683869836316194948\cdots.
\end{aligned} 
\end{equation}
The appearance of the function $f(s)=2s/\cosh s$
is expected from the semi-classical limit, as we will see in section \ref{sec:semi}.

In general, we expect that the partition function is written as
an infinite product
\begin{equation}
\begin{aligned}
 Z(\bt)=\prod_{i=1}^\infty\left[1+\frac{\bt^2}{t_i(q)^2}\right],
\end{aligned} 
\end{equation}
where $t_i(q)$ is the $i$-th zero of $Z(\bt)$ along the imaginary $\bt$ axis.
The first zero $t_1(q)$ is given by \eqref{eq:pole-t}.

\section{Pad\'{e} approximation}\label{sec:pade}
\begin{figure}[t]
\centering
\includegraphics
[width=0.8\linewidth]{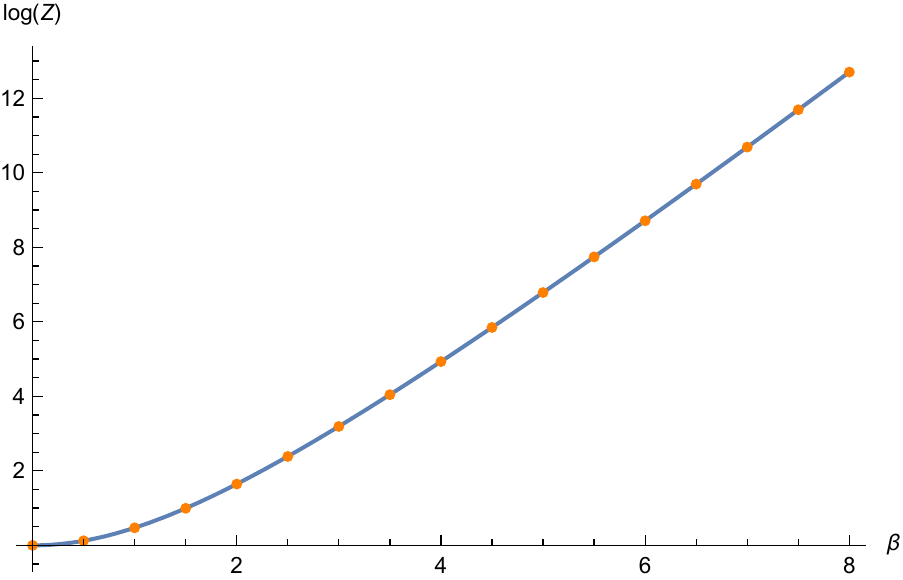}
  \caption{
Plot 
of free energy $\log Z(\bt)$ for $q=0.2$.
The blue solid curve represents the Pad\'{e} approximation
of the small $\bt$ expansion \eqref{eq:pade} 
with $n_{\text{max}}=100$, while the orange dots are the exact result
in \eqref{eq:Z-int}.
}
  \label{fig:pade}
\end{figure}

Lets us go back to the real $\bt$ case.
Using our data of $\{f_n(q)\}_{1\leq n\leq n_{\text{max}}}$
with $ n_{\text{max}}=100$, we can improve the
approximation of the small $\bt$ expansion in \eqref{eq:free2}
by the (diagonal) Pad\'{e} approximation
\begin{equation}
\begin{aligned}
 \log Z(\bt)&\approx \sum_{n=1}^{n_{\text{max}}}(-1)^{n-1}(1-q)^{n-1}
\bt^{2n}f_n(q)\\
&\approx \frac{\sum_{i=0}^{n_{\text{max}}}b_i\bt^i}{\sum_{i=0}^{n_{\text{max}}}a_i\bt^i},
\end{aligned} 
\label{eq:pade}
\end{equation}
where the second line is the Pad\'{e} approximant 
constructed in such a way that the small $\bt$ 
expansion of the second line agrees with the first line
up to $\cO(\bt^{2n_{\text{max}}})$.
We can compare this Pad\'{e} approximation 
with the exact result in \eqref{eq:Z-int}.
In Figure \ref{fig:pade}, we show the plot of free energy $\log Z(\bt)$ for $q=0.2$
in the Pad\'{e} approximation (blue solid curve) and the exact
result in \eqref{eq:Z-int} evaluated numerically (orange dots). 
As we can see from Figure \ref{fig:pade}, the 
Pad\'{e} approximation of the small $\bt$ expansion exhibits
a good agreement with the exact result even at large $\bt$.

This agreement suggests that there is no phase transition
between the high temperature regime and the low temperature regime and
they are smoothly connected.
This is in contrast to the situation for the coupled SYK model
considered in \cite{Maldacena:2018lmt}
where the high and the low temperature phases are separated by the Hawking-Page
transition.  Our result suggests that
there is no Hawking-Page transition for the partition function
$Z(\bt)$ of DSSYK.
 
\section{Semi-classical expansion}\label{sec:semi}
As discussed in \cite{Goel:2023svz},
the $\th$-integral in \eqref{eq:Z-int} can be evaluated by the 
saddle-point approximation in the semi-classical small $\la$ regime.
To define the systematic small $\la$ expansion,
we have to rescale $\bt\to \bt\la^{-\hf}$
\begin{equation}
\begin{aligned}
 Z(\bt\la^{-\hf})=
\int_0^\pi\frac{d\th}{2\pi}(q;q)_\infty (e^{2\ri\th};q)_\infty
(e^{-2\ri\th};q)_\infty \exp\left(\frac{2\bt\cos\th}{\rt{\la(1-q)}}\right).
\end{aligned} 
\end{equation}
As shown in \cite{Okuyama:2023bch}, in the small
$\la$ limit the measure factor
is expanded as
\begin{equation}
\begin{aligned}
 (q;q)_\infty (e^{2\ri\th};q)_\infty
(e^{-2\ri\th};q)_\infty=\rt{\frac{2\pi}{\la}}
\exp\left[\frac{\la}{8}-\frac{2}{\la}\left(\th-\frac{\pi}{2}\right)^2
+\log(2\sin\th)\right].
\end{aligned} 
\end{equation}
Note that there are no corrections higher than $\cO(\la^2)$ in the measure factor.
Now, we can evaluate the $\th$-integral by the saddle point approximation.
At the leading order of the small $\la$ expansion, 
the saddle point equation is given by
\begin{equation}
\begin{aligned}
 \frac{\del}{\del\th}\left[-2\left(\th-\frac{\pi}{2}\right)^2
+2\bt\cos\th\right]=0,
\end{aligned} 
\end{equation}
and the saddle point value of $\th$ is given by
\begin{equation}
\begin{aligned}
 \th=\frac{\pi}{2}-u
\end{aligned} 
\end{equation}
where $u$ is related to $\bt$ as\footnote{
Our $u$ is related to $v$ in \cite{Maldacena:2016hyu} by $u=\pi v/2$.
}
\begin{equation}
\begin{aligned}
 \bt=\frac{2u}{\cos u}.
\end{aligned} 
\label{eq:bt-u}
\end{equation}
One can systematically improve the approximation by expanding the 
integrand around the saddle point
\begin{equation}
\begin{aligned}
 \th=\frac{\pi}{2}-u+\rt{\la}x
\end{aligned} 
\end{equation}
and perform the Gaussian integral over the fluctuation $x$.
In this way we find the semi-classical, small $\la$
expansion of the free energy
\begin{equation}
\begin{aligned}
 \log Z(\bt\la^{-\hf})=\sum_{k=0}^\infty\la^{k-1}\cF_k(u).
\end{aligned} 
\end{equation}
The first few terms of $\cF_k(u)$ read
\begin{equation}
\begin{aligned}
 \cF_0(u)&=-2u^2+4u\tan u,\\
\cF_1(u)&=u\tan u+\log(\cos u)-\hf\log(1+u\tan u),\\
\cF_2(u)&=\frac{4 u^4 \tan ^4u+12 u^4 \tan ^2u-12 u^3 \tan ^3u+12 u^3 \tan u+5 u^2-21 u^2 \tan ^2u-5 u
   \tan u}{96 (u \tan u+1)^3},\\
\cF_3(u)&=\frac{1}{768 (u \tan u+1)^6}
\Bigl[-8 u^7 \tan ^7u-16 u^7 \tan ^5u-8 u^7 \tan ^3u-24 u^6 \tan ^6u\\
&\quad -24 u^6 \tan ^4u-42 u^5
   \tan ^5u-56 u^5 \tan ^3u-6 u^5 \tan u+u^4-25 u^4 \tan ^4u-56 u^4 \tan ^2u\\
&\quad +27 u^3 \tan^3u-15 u^3 \tan u-7 u^2+33 u^2 \tan ^2u+7 u \tan u\Bigr].
\end{aligned} 
\label{eq:cFk}
\end{equation}
We are interested in the small $\bt$ regime, 
which corresponds to the small 
$u$ regime. The small $u$ expansion of $\cF_k(u)$ 
in \eqref{eq:cFk} is easily found as
\begin{equation}
\begin{aligned}
 \cF_0(u)&=2 u^2+\frac{4 u^4}{3}+\frac{8 u^6}{15}+\frac{68 u^8}{315}+\frac{248 u^{10}}{2835}+\frac{5528
   u^{12}}{155925}+\frac{87376 u^{14}}{6081075}+\frac{3718276 u^{16}}{638512875}+\cdots,\\
\cF_1(u)&=\frac{u^4}{3}+\frac{2 u^6}{45}+\frac{23 u^8}{315}+\frac{34 u^{10}}{14175}+\frac{8014
   u^{12}}{467775}-\frac{20236 u^{14}}{8513505}+\frac{273653 u^{16}}{58046625}
+\cdots,\\
\cF_2(u)&=-\frac{u^4}{9}+\frac{2 u^6}{9}-\frac{55 u^8}{189}+\frac{2854 u^{10}}{8505}-\frac{161786
   u^{12}}{467775}+\frac{6119348 u^{14}}{18243225}-\frac{4744967 u^{16}}{15324309}
+\cdots,\\
\cF_3(u)&=\frac{u^4}{36}-\frac{17 u^6}{90}+\frac{299 u^8}{540}-\frac{19979 u^{10}}{17010}+\frac{1900147
   u^{12}}{935550}-\frac{55729109 u^{14}}{18243225}+\frac{31769442277
   u^{16}}{7662154500}+\cdots.
\end{aligned} 
\label{eq:cFk-exp}
\end{equation}
On the other hand, we can find the small
$\la$ expansion of the free energy directly
from \eqref{eq:free2}
\begin{equation}
\begin{aligned}
 \log Z(\bt\la^{-\hf})&=\sum_{n=0}^\infty
(-1)^{n-1}\bt^{2n}\la^{-n}(1-q)^{n-1}f_n(q)\\
&=\sum_{k=0}^\infty \la^{k-1}F_k(\bt).
\end{aligned}
\label{eq:free3} 
\end{equation}
Thanks to the presence of 
the factor $(1-q)^{n-1}$ in the expansion of free energy
\eqref{eq:free2}, one can take a well-defined semi-classical limit after rescaling
$\bt\to\bt\la^{-\hf}$. 
The first few terms of $F_k(\bt)$
in \eqref{eq:free3}  read
\begin{equation}
\begin{aligned}
 F_0(\bt)&=
\frac{\bt^2}{2}-\frac{\bt^4}{24}+\frac{\bt^6}{120}-\frac{23 \bt^8}{10080}+\frac{67 \bt^{10}}{90720}-\frac{5297
   \bt^{12}}{19958400}+\frac{997 \bt^{14}}{9729720}-\frac{26618 \bt^{16}}{638512875}
+\cdots,\\
F_1(\bt)&=\frac{\bt^4}{48}-\frac{7 \bt^6}{720}+\frac{61 \bt^8}{13440}-\frac{247 \bt^{10}}{113400}+\frac{64091
   \bt^{12}}{59875200}-\frac{8849 \bt^{14}}{16511040}+\frac{712196791
   \bt^{16}}{2615348736000}+\cdots,\\
F_2(\bt)&=-\frac{\bt^4}{144}+\frac{\bt^6}{144}-\frac{43 \bt^8}{8064}+\frac{289 \bt^{10}}{77760}-\frac{590147
   \bt^{12}}{239500800}+\frac{822299 \bt^{14}}{518918400}-\frac{156670559
   \bt^{16}}{156920924160}+\cdots,\\
F_3(\bt)&=
\frac{\bt^4}{576}-\frac{11 \bt^6}{2880}+\frac{11 \bt^8}{2304}-\frac{10417 \bt^{10}}{2177280}+\frac{254201
   \bt^{12}}{59875200}-\frac{2170187 \bt^{14}}{622702080}+\frac{21258243901
   \bt^{16}}{7846046208000}+\cdots.
\end{aligned}
\label{eq:Fkbt} 
\end{equation}
The small $u$ expansion of $\bt$ in \eqref{eq:bt-u} is given by
\begin{equation}
\begin{aligned}
 \bt=\frac{2u}{\cos u}=2u\sum_{n=0}^\infty \frac{(-1)^nE_{2n}u^{2n}}{(2n)!},
\end{aligned}
\label{eq:bt-expand} 
\end{equation}
where $E_{2n}$ denotes the Euler number. Plugging
\eqref{eq:bt-expand} into $F_k(\bt)$ in 
\eqref{eq:Fkbt}, we find that the small $u$ expansion of 
$F_k(\bt)$ agrees with
$\cF_k(u)$ in \eqref{eq:cFk-exp}
\begin{equation}
\begin{aligned}
 F_k\left(\frac{2u}{\cos u}\right)=\cF_k(u).
\end{aligned} 
\label{eq:Fk-agreee}
\end{equation}

To summarize, we find that the semi-classical small $\la$
expansion obtained from
the saddle point approximation of the exact result is 
consistent with the small $\bt$ expansion 
of free energy in \eqref{eq:free2} after rescaling
$\bt\to\bt\la^{-\hf}$.
In other words, there is no order-of-limit problem between the small
$\la$ expansion and the small $\bt$ expansion
and we find the agreement of $\cF_k(u)$ and $F_k(\bt)$ in \eqref{eq:Fk-agreee}.

Finally, we comment on the interpretation of $A(1)$ 
we found in section \ref{sec:zero}.
The analytic continuation of $\bt$ to the imaginary value
$\bt=\ri t$ corresponds to the analytic continuation of 
$u$ to the imaginary value $u=\ri s$.
Then the relation \eqref{eq:bt-u} becomes
\begin{equation}
\begin{aligned}
 t=\frac{2s}{\cosh s}.
\end{aligned} 
\end{equation}
As we increase $t$ from zero, this relation 
ceases to have a solution above the maximum of the
right hand side.
Thus, it is natural that $A(1)=\max\{2s/\cosh s\}$ sets the value of 
the radius of convergence when $q=1$,
if we regard the rescaled combination 
$\bt\rt{1-q}$ as the expansion parameter.

\section{Conclusions and outlook}\label{sec:conclusion}
In this paper, we have studied the high temperature expansion of the free
energy of DSSYK.
We found that the radius of convergence of this expansion is determined by the first
zero of the partition function along the imaginary $\bt$-axis.
We have also computed the first few terms
of the semi-classical expansion of free energy and found that they
are consistent with the small $\bt$ expansion at finite $q$.

There are many interesting open questions.
From the analysis of the Pad\'{e} approximation of the small $\bt$
expansion and the numerical evaluation of the exact $Z(\bt)$ in
\eqref{eq:Z-int}, we find that there is no Hawking-Page transition for the partition
function of DSSYK and the high and the low temperature regimes are smoothly
connected.
On the other hand, it is speculated that 
the UV completion of the SYK model
involves stringy degrees of freedom \cite{Maldacena:2016hyu,Goel:2021wim}.
Our result suggests that there is no Hagedorn
growth of the degrees of freedom in DSSYK at high energy.
In fact, the eigenvalue spectrum of the transfer matrix
$T$ in \eqref{eq:T-mat} is bounded from above.
It would be interesting to understand the stringy degrees of freedom in
DSSYK better.

In general, we are still lacking a clear understanding of the bulk 
spacetime picture of DSSYK.
In particular, we would like to understand the
bulk dual of DSSYK at finite $\la$, away from the semi-classical limit.
As discussed in \cite{Avdoshkin:2019trj},
the moment $\bra0|T^{2n}|0\ket$ can be expanded as a sum over the Dyck paths
and the limit shape of the Dyck paths are obtained for the case of
the ordinary (not $q$-deformed) oscillators \cite{Avdoshkin:2019trj}.
It would be interesting to find the limit shape of the Dyck paths for the 
$q$-oscillator case, 
which might be related to the Brownian motion of the boundary particle
discussed in \cite{Berkooz:2022mfk}.
We leave this as an interesting future problem.

\acknowledgments
The author would like to thank Matthieu Josuat-Verg\'{e}s for correspondence.
This work was supported in part by 
JSPS Grant-in-Aid for Transformative Research Areas (A) 
``Extreme Universe'' 21H05187 and JSPS KAKENHI 22K03594.
%%%%%%%%%%%%%%%
\bibliography{paper}
\bibliographystyle{utphys}

\end{document}